Statistical inference for Axiom A attractors

M. LuValle

Rutgers University

**Abstract**

From the climate system to the effect of the internet on society, chaotic systems appear to have a significant role in our future. Here a method of statistical learning for a class of chaotic systems is described along with underlying theory that can be used not only for predicting those systems a short time ahead, but also as a basis for statistical inference about their dynamics. The method is applied to prediction of 3 different systems. The statistical inference aspect can be applied to explore and enhance computer models of such systems which in turn can provide feedback for even better prediction and more precise inference.

## Introduction

Chaotic systems can be characterized as deterministic dynamic systems for which prediction becomes exponentially more difficult as the delay between the observations and the prediction increases. Several problems currently facing the human race might be characterized as being driven by such dynamic systems, ranging from the relationship between internet availability and the increase in mass shootings in the United States, to the increase in category 4 and 5 hurricanes in the Atlantic basin over the last 100 years. Although we have developed tools that may allow us to quantitatively follow these events (e.g. using the GLOVE [1] paradigm to vectorize social media interaction and using combinations of sophisticated satellite monitoring and ground measurements of weather and climate) the ability to predict, and to make statistical inferences to test hypothesis about these systems is still developing.

Two of the difficulties in developing a theory of statistical inference for real dynamic systems are

1) There are not yet tools for characterizing the statistical error induced by the chaotic dynamics particularly when modeled by low dimensional models.
2) None of the real systems are actually in a stationary state, with both impulses coming in to the system from the outside and fundamental parameters tuning the system changing over time.

For the latter issue, with slow changes of the parameters, if the attractor changes continuously (and even better with continuous derivatives) with the parameters, then an accurate computer model could be used to approximate the statistics and develop statistical inference for the system. There is strong indication that high dimensional chaotic systems in general have at least the continuous property [2,3], as do Axiom A systems [4]. Under these assumptions fairly accurate predictions for regional seasonal precipitation using linear response [5,6] approximations have been obtained several seasons ahead [7,8]. However, to judge if any error is due to the inherent unpredictability, bad approximation using a linear model, or a fundamental error in the computer model, a rigorous approach to statistical estimation (and inference) beyond pure linear response is necessary.

The “chaotic hypothesis” [9] states that “a many particle system in a stationary state may be regarded as a smooth dynamical system with a transient Axiom A global attractor for purpose of computation of its microscopic properties…". In other words most real high dimensional chaotic systems might be expected to behave like axiom A systems.

Further axiom A systems [10] have natural probability measures [11,12] and ergodic properties, that along with the topological properties of general chaotic systems [13] lend themselves to an asymptotically consistent framework for statistical inference.

In the second section the topological theory of observations from chaotic systems [13] and the ergodic theory of Axiom A systems [10-12] is described and developed in the direction of making it a framework for asymptotically consistent statistical inference proving a sequence of theorems ending with two conjectures. The theory starts with consistent prediction from precise observation of a stationary axiom A system and proceeds to theory estimating the convolution of the SRB measure induced by combining predictors from several projections. Then conjectures are made concerning a resampling idea for statistical characterization of predictions, estimation with both small external perturbations, and non-stationarity, and estimation beyond the inverse of the maximum Lyapunov exponent. The third section shows application of the approach developed in the theorems and conjectures, first to prediction along the Lorentz attractor, in which a fourth conjecture falls out of the experimental data, then to the stock market, and finally to a sequence of precipitation measurements at Volcano national park in Hawaii. In the first two cases only the predictions are calculated, in the latter case the method for estimating the deviations from the predictions described in the theory section is applied to generate empirical estimates of prediction bounds for the system, and the resampling method is used to bound the estimated correlation between predictions and the true values. The discussion outlines some ideas for using neighborhoods of computer models to build approximate statistical tests of whether a real system is well captured by the computer models.

## Results

### Mathematical Development

For mathematical purposes Axiom A attractors, with twice differentiable paths are relatively easy to work with. They decompose into 3 mutually exclusive sets of manifolds, contracting, expanding and stationary. A method of estimating a manifold with iid (independent and identically distributed) errors in high dimensional space in a way that is asymptotically optimal (for iid errors) has been described [14]. In that paper [ibid] a generalized cross validation procedure is used to pick a bandwidth for a kernel linear regression estimator of the local surface. The dimension going into the kernel is chosen by a maximum likelihood estimate of the dimensionality based on Poisson approximation [15]. For the purpose of predicting new observations that approach will provide local accuracy to the first order, however we apply it in the context of multiview embedding [16] where multiple embeddings [13] of the attractor using delay maps [ibid, defined later] with a set response variable are combined to make an prediction. In particular a predictor will be built for each delay map, at each point to be predicted. The embedding theorems [ 13] state that almost all delay maps with dimension large relative to the box dimension of the attractor are almost surely diffeomorphic with respect to the original attractor. Unfortunately, in real world systems suspected of chaotic behavior, such as the climate system, and the stock market, the dimension of the attractor is unknown, difficult to identify, and probably high. On a more positive note it has been shown [17] that even projections onto single dimensions using nearest neighbor estimates can result in non-trivial prediction. Instead of setting up the full mechanics for dimension and variable selection, in our examples we define the dimension and neighborhoods crudely on training data.

To begin, consider a stationary chaotic system (i.e. the tuning parameters of the system are stationary). We will start with a simple case, where the attractor of the chaotic system is of known dimension so the answer is clear from prior work [10,11,13].

To set up a framework for statistical inference, we need an underlying probability distribution for the system, and observable quantities derived from the system. For the probability distribution we use the SRB measure [10-12]. We use definition 3 from Young[12]: If f is a C2 diffeomorphism on an attractor A and suppose that there is a unique SRB measure $\mu$, then there is a set $V \subset U$ (the latter the basin of the attractor) having full Lebesgue measure (the measure of the set U\V is 0) such that for every $\varphi: U \to R, \frac{1}{n}\sum_{i=0}^{n-1} \varphi\left(f^{i}(x)\right) \to \int \varphi d\mu$ , for every $x \in V$. This gives us our probability measure with a purely frequentist definition, defined only based on the ergodicity (e.g. let $\varphi$ .be the indicator function for a subset of V) of Axiom A attractors [12].

I assume for the remainder of the paper that the chaotic system can be defined by a system of differential equations, then the boundary conditions and parameters of the system define the parameters of that probability distribution, say $\theta$.

For the observable quantities, we review some of the results of Sauer et al [13] that allow the first synthesis that results in asymptotically consistent predictions of extremes of a chaotic system. We start with Sauer et al.'s [13] theorem 2.7. The precise statement of the result requires a few mathematical definitions. The box dimension of a compact set A in an n dimensional Euclidean space is $box\ dim(A) = \lim_{\varepsilon \to 0} \left(\frac{log(N_\varepsilon)}{-log(\varepsilon)}\right)$ where $N_\varepsilon$ is the number of cubes with side $\varepsilon$that it takes to cover A. Again following Sauer (ibid) a general delay coordinate map takes the form $F(x) = \left(h_1(x), \ldots, h_1\left(g^{p_1-1}(x)\right), \ldots, h_j(x), \ldots, h_j\left(g^{p_j-1}(x)\right)\right)$, where $h_i(x)$ is the ith coordinate of a measurement of a term on a chaotic attractor, and $g^k(x)$is k fold application of a diffeomorphism or flow on A. Let $p = p_1 + \ldots + p_j$. Theorem 2.7 [13] now states, given some restrictions on periodic orbits, that if $p > 2box\ dim(A)$then F is both 1-1 on A and an *immersion* (derivatives are 1-1) on every smooth manifold C contained in A.

A concrete example of a delay map in studying monthly precipitation follows: there are k weather stations close to given position, t(0) is the time to predict, and weather station 1 is where we would want to predict. Define Pi(t(j)) as precipitation at station i at time t(j), Ti(tj) as the corresponding temperature at station i. Then to create a delay coordinate map we could start with the vector (P1(t(0), P1(t(-5),T1(t(-5),P2(t(-5)),Pk(t(-7)) for a total of 4 measurements. This map would then include stepping each measurement back 1 in time(P1(t(-1), P1(t(-6),T1(t(-6),P2(t(-6)),Pk(t(-8)), etc., for a sequence of multivariate measurements in time and space, defined as a delay map. Our target prediction is P1(t) where we are prediction 5 units ahead in time.

Consider the following construction: Take a delay coordinate map where we identify the furthest coordinate forward in time as a statistically dependent variable (y) which we want to predict. The remainder are $p > 2box\ dim(A)$ independent variables. Now apply nearest neighbor regression (i.e., identify nearest m neighbor points with respect to X, and build a linear model of y on these p variables to predict y). Strange attractors are confined to compact sets. By the ergodicity of the SRB measure, and since the mapping is an immersion the local regression plane is converging to the tangent plane of the manifold. The nearest neighbor multiple regression provides an extrapolation correct to the 1st order. The ergodic nature of the chaotic system[10-12] and topological nature of delay maps [13] thus ensures asymptotically consistent prediction for extreme values for example taking an m nearest neighbor regression using p independent variables with m=o(number of observations) and p > 2*boxdim(A).

This proves

**Theorem 1: With direct observation of an axiom A system with known box dimension d, applying Bickel's method with p>2d results in consistent prediction of the system.**

Now let us consider the case where boxdim(A) is unknown. The obvious first step is to let p grow as o(m) in the above scenario, but we would like to know that along the way before p> 2*boxdim(A) we can get useful predictions. To this point there is further theory in Sauer (theorem 2.10) that demonstrates very small probability is assigned to the region where non-immersion occurs for boxdim<p<2*boxdim(A).

It turns out we can bound the predictability of the tangent planes a little bit better based on a further result of Sauer et al [14], and some extension [18].

Multiview embedding [16] has shown that using several p variable regressions can result in better accuracy. It can be argued [18] that disjoint sampling is in fact a more efficient way to converge to accurate prediction when using Multiview embedding.

To develop this, we turn to Sauer et al, to theorem 2.10[13]. One more definition is required for stating theorem 2.10 [13]. The $\delta$ -distant self intersection set is defined as $\Sigma(F,\delta) = \{x \in A: F(x) = F(y) \text{ for some } y \in A, |x-y| > \delta\}$. Then theorem 2.10 states that if A is compact in $R^k$with $box\, dim(A) = d$, then if $p \leq 2d$, $\Sigma(F, D)$ has lower box counting dimension at most $2d - p$, and $F$is an immersion on each compact subset C of an m-manifold contained in A except on a subset of C of dimension at most 2m-n-1.

Now define$F_1, F_2$ to be strictly distinct if the values of their respective $h_i$do not overlap. Then:

**Lemma 1: For almost all $F_1, F_2$ strictly distinct with dimensions $p_1, p_2$, if $d < p_1 < p_2 \leq 2d$, then $\Sigma(F_1,\delta) \cap \Sigma(F_2,\delta) = \varphi$.**

PROOF: Suppose there is a set of $F_i, F_j$ strictly distinct, with positive probability such that $\varphi \neq \Sigma(F_i,\delta) \cap \Sigma(F_j,\delta)$, for $i, j$. Then it would be possible to construct a set of product maps $(F_i, F_j)$ with positive probability for which $\varphi \neq \Sigma\left((F_i, F_j),\delta\right)$, even though the dimension of the product map is $p^{i,j} = p + p_j > 2d$, where $p^{i,j}$is the dimension of $(F_i, F_j)$contradicting theorem 2.7 of Sauer (ibid).

Note that if $F_i, F_j$ are distinct, but not strictly distinct, the outcome then depends on the overlap. If the number of overlapping coordinates is such the $p^{i,j} > 2d$ the result holds. If $p^{i,j} \leq 2d$, the null set may be common, but cannot be larger than $2d - p^{i,j}$.

Suppose we wish to model $x$, a term in $F_i, i = 1,2$and suppose that we have a consistent estimator of $E(x|F_i)$. Suppose $d < p_i < 2d$ . Let$\Sigma_i = \Sigma(F_i,\delta), i = 1,2$. Then we can decompose the attractor into 4 mutually exclusive sets, $S_1 = \{\Sigma_1 \cap \Sigma_2\}, S_2 = \{\Sigma_1 \cap \bar{\Sigma}_2\}, S_3 = \{\bar{\Sigma}_1 \cap \Sigma_2\}, S_4 = \{(\bar{\Sigma}_1 \cap \bar{\Sigma}_2)\}$. Suppose we use our hypothetical consistent estimator on each $F_i$ and look at the pairs of prediction based on each.

Exhibit 1

$$E_4\left(\left(E(x|F_1), E(x|F_2)\right)\right) = (x, x)$$

$$E_3\left(\left(\left(E(x|F_1), E(x|F_2)\right)\right)\right) = \left(x, E_3(x, F_2)\right)$$

$$E_2\left(\left(E(x|F_1), E(x|F_2)\right)\right) = (E_2(x|F_1), x)$$

$$E_1\left(E(x|F_1), E(x|F_2)\right) = \left(E_1(x|F_1), E_1(x|F_2)\right)$$

So, the true value shows up in regions $S_2, S_3, S_4$. Taking advantage of corollary 1, making $F_1$, and $F_2$ strictly distinct, $S_1 = \varphi$ so the true value of the attractor is at least one of the predictions. With a large number of mutually disjoint delay maps of this size at any given x only one can have $\Sigma_i$ covering that x.

The above disjoint system works only for boxdim(A)<p<2*boxdim(A).

This leads us to:

**Theorem 2: if d<P<2d and we have a set $\boldsymbol{\Theta}$ disjoint $\boldsymbol{F_i}$ of dimension P, With direct observation of an axiom A system applying this multiview embedding version of Bickel's method removing at each x a small percentage of the outlying predictions results in consistent prediction of the system.**

However Garland and Bradley [8]have shown that useful predictions can arise even with p ( the number of independent variables)=1 in higher dimensional systems.

To understand the latter case, start with a full embedding in which the p independent variables are a subset of those needed for a full reconstruction, say p1 such variables. The marginal distribution of y, the dependent variable, given the p independent variables is the integral across the p1-p variables with respect to the SRB measure. By the boundedness of the functions and dominated convergence and the nature of V in the definition of SRB measure for an axiom A attractor and Leibniz's rule, the integral is differentiable. Linear regression in the ball converges to the hyperplane closest in the least squares case to the mean of the manifolds in the ball, which will itself be a manifold. This mean manifold is differentiable through Leibniz's rule with a tangent plane from its differentiability. As the ball contracts the average manifold approaches the tangent plane, giving us Lemma 2.

**Lemma 2: In the case that p<d, local linear regression converges to the mean of the manifolds contained in the cylinder set of the ball. By the conditions on Axiom A systems, this mean is a differentiable manifold, so the local linear regression converges to the tangent plane as the ball collapses as the sample size increases.**

In this case, if p is not a full embedding, rather than predicting extremes directly, the empirical distribution of residuals around the regression plane provides a basis for an asymptotically consistent probabilistic prediction of the extremes. To enable that prediction, for each delay map, consider dividing the data set into two portions the first qn to predict the last (1-q)n (0<q<1) using the Bickel approach. Let k the number of nearest neighbors increase, but assume k=o(n), and for each predicted point in the last (1-q)n points retain the residual from the nearest neighbor in the first qn points to its delay map. For any neighborhood of this predicted point, this gives a conditional distribution of residuals that converges to the induced conditional SRB measure of the attractor around the collapsing induced by overly small p. Note that in the case of symmetric measurement errors, this will also induce a consistent estimate of the error distribution convoluted with the induced SRB measure around the predicted point.

Note that in the case of multiview embedding with p< 2d, the predictive distribution is a convolution of induced conditional SRB measures with the sampling induced distribution of cylinder sets of the partition. This convolution will still provide an asymptotically consistent probabilistic bound, based on the uncertainty induced by both the sampling process and induced conditional SRB measure.

**THEOREM 3: Suppose d is unknown, then letting k increase, k=o(n), and p increase p=o(k), and finally retaining the single nearest neighbor residual to use for building a predictive distribution around the estimate:**

**a) The prediction converges as n, k and p increases**

b) **Even for fixed p, the conditional predictive distribution based on the single residual for the nearest neighbor converges to a predictive distribution, both in the case of perfect measurement and the case of a symmetric 0 mean measurement error distribution.**

The predicted values are converging to the true value by theorems 1-3. Further the conditional predictive distributions are converging correctly under our assumptions. Each partition into disjoint delay maps induces several estimates of the prediction surface. At each point to be predicted, these estimates are combined by trimmed mean to yield a prediction at each point. The compactness of the sample space, and the fact that the nearest neighbor linear regression estimates use a number of nearest neighbors which is o(n) where n is the number of observations implies the points going into the estimate are getting farther apart in time. Adding the assumption that our systems are rapidly mixing. Then the correlation is approaching 0 and the variance of a mean of several such surfaces shall converge to an average variance divided by the number of points. Independently sampled partitions have independently sampled delay maps. Assuming a central limit theorem for the axiom A flow [19], then the estimates are converging to statistically independent estimates so the random partition sampling could be conjectured to be asymptotically acting like bootstrap sampling, and empirical distributions produced by such random samples should provide confidence bounds for estimates. This leads to:

**Conjecture 1**: Approximate inference for summary statistics of stationary s axiom A systems. As n increases for window size=o(n), where window size is also increasing, estimates based on a nearest neighbor strategy predictions for distinct measurements act as if they are independent, so correlations and inference for correlations and other summary statistics can be calculated using resamples based on different random partitions of the delay map space. This idea is applied in example 3 with some support provided by the data analysis. Rigorous theoretical support might be derived from the similarity of these estimates with U statistics.

**Conjecture 2**: Suppose now that in addition to the scenario for theorem 3, there are intermittent innovations impinging at random on the dynamic system, since the space decomposes into direct sums of contracting, and expanding manifolds, the innovation can be similarly decomposed. The portion in the contracting manifold moves towards the manifold, while that in the expanding manifold simply redefines the point at the time of the innovation. The fluctuation dissipation results [11,12] imply a locally linear relationship in the relaxation of the response toward the tangent plane of theorems 1-3. Similarly a slow change in tuning parameters would have a similar response as long as the attractor varies smoothly with changes in the tuning parameters, which for high dimensional systems seems to hold [13,14].Thus the Bickel estimate in both cases has a tangent plane (or a time averaged tangent plane, approximating the tangent plane at the time being predicted) to converge to. In both the cases of random innovations, and in the case of sufficiently slow changes in the tuning parameters, the method of theorem 3 would be expected to result in consistent estimates both for prediction and the conditional predictive distribution.

**Conjecture 3:** Measurement error limits the ability to extrapolate very far forward in a chaotic system (along with external influences over time). However, the conditional SRB measure does not immediately inflate to the stationary SRB measure as one passes the inverse of the largest Lyapunov exponent. Thus this method should have the potential to provide useful extrapolation beyond the usual range even though it will not be precise.

## Applications

In this section the method is applied to 3 systems to predict chaotic behavior. The basic window is sqrt(n) with multipliers checked using cross validation to choose the multiplier The classic Lorentz attractor, 75 years of stock market data for three Dow Jones averages, and 60 years of monthly precipitation data from a weather station in Volcano national park in Hawaii. In the first two systems only the predictions are used, in the last, prediction bounds are shown based on the theorem 3. And we examine the correlation of predictions vs observations in light of Conjecture 1 showing that the variance of the estimated correlation across independendent random partitions of delay maps behaves as if there is 0 correlation, and appear asymptotically normal

In the first case a naïve purely linear model is compared to the Bickel method for a number of values of P and neighborhood sizes. In the second, predictions of the next days direction for stocks is used to determine whether to sell out at the close of market, stay in, or buy back in, and in the third predictions and prediction bounds are provided for the deviation of the monthly precipitation from the historical monthly average precipitation 1 year ahead.

### Application 1) Lorentz attractor

Figures 1-3 show the progression for multiview linear regressions (on the left) compared to the multiview local linear regressions on the right, as the dimension of the x matrix in the delay map changes from 2 to 5 to 25(so the actual delay map ranges in dimension are 3, 6 and 26 as the term being predicted is part of the delay map). The X axis in each plot is the prediction from the model, the y axis is the observed value, The predictions are 20 Time steps ahead. We see the local linear model performs well, but in fact continues to improve even beyond the delay map dimension of 5 (>twice the box dimension of ~2.1) . At dimension 25 in figure 3 and we see that the linear model has become indistinguishable from the local linear model at the level of accuracy we are calculating.

**Figure 1. P=2**

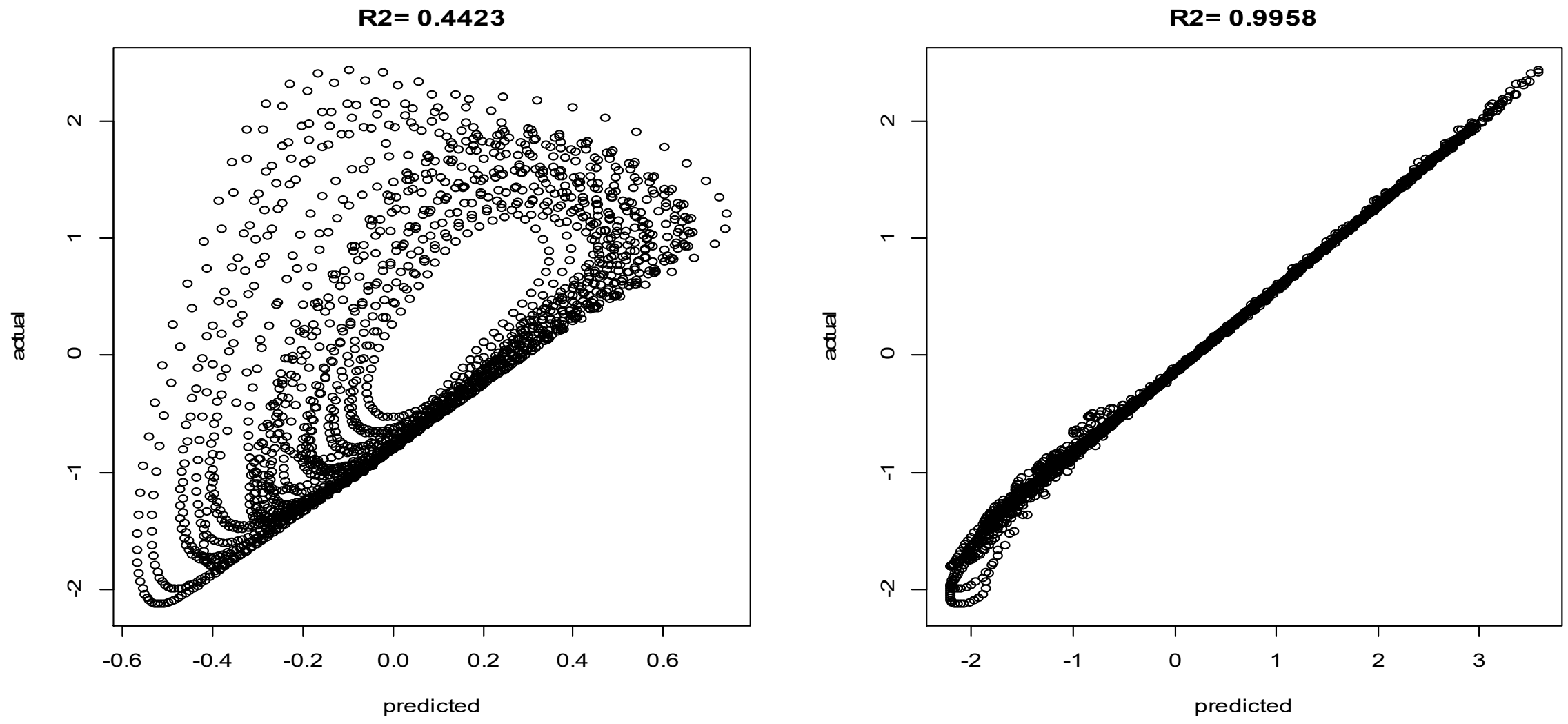

*Caption Figure 1: The plot on the left is the predicted vs observed value using the average linear fit to 3 dimensional delay maps with 2 independent dimensions predicting the third, extracting the 2 from a potential 30 dimensional delay map. The fit on the right is the average local linear fit using the same partition with a nearest neighbor size of 10*sqrt of n. The training is from the first 4000 steps in a 5000 step depiction of the Lorenz attractor.*

**Figure 2.** P=5

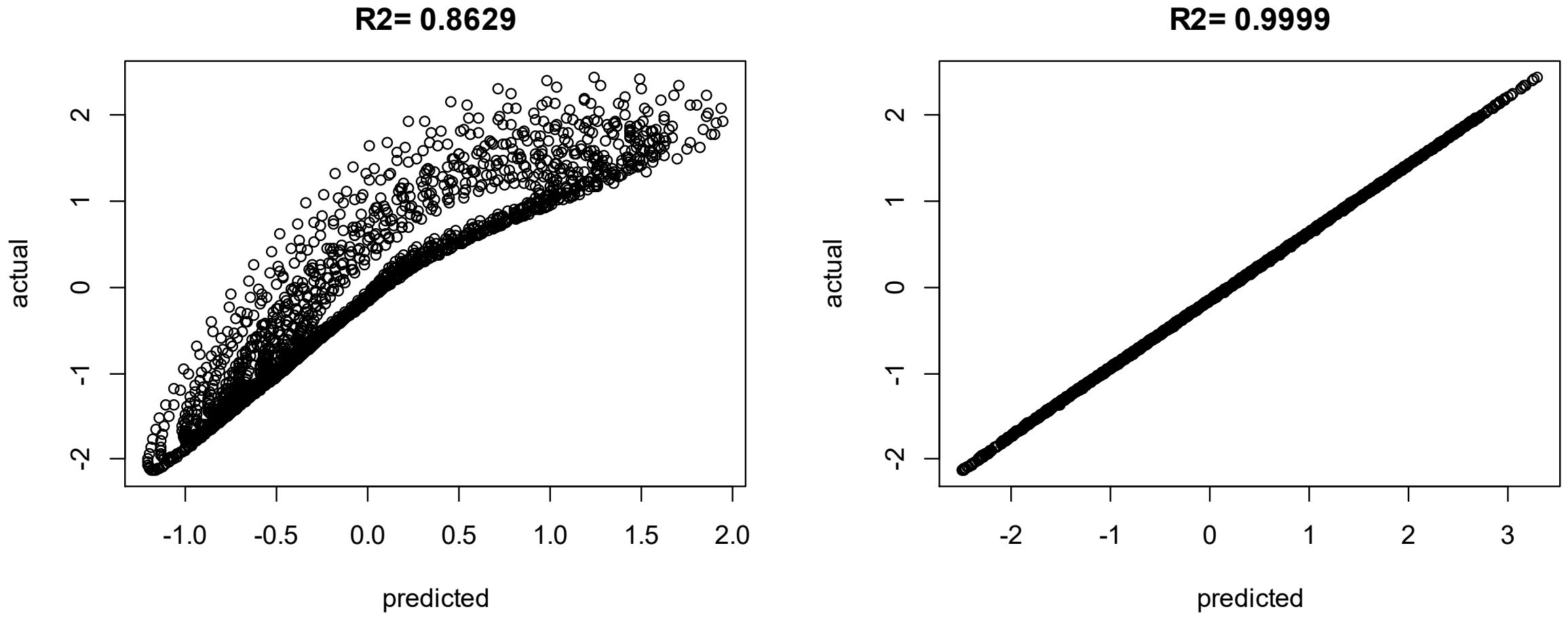

*Caption Figure 2: The plot on the left is the predicted vs observed value using the average linear fit to 6 dimensional delay maps with 5 independent dimensions predicting the third, extracting the 5 from a potential 30 dimensional delay map. The fit on the right is the average local linear fit using the same partition with a nearest neighbor size of 10*sqrt of n. The training is from the first 4000 steps in a 5000 step depiction of the Lorenz attractor.*

**Figure 3. P=25**

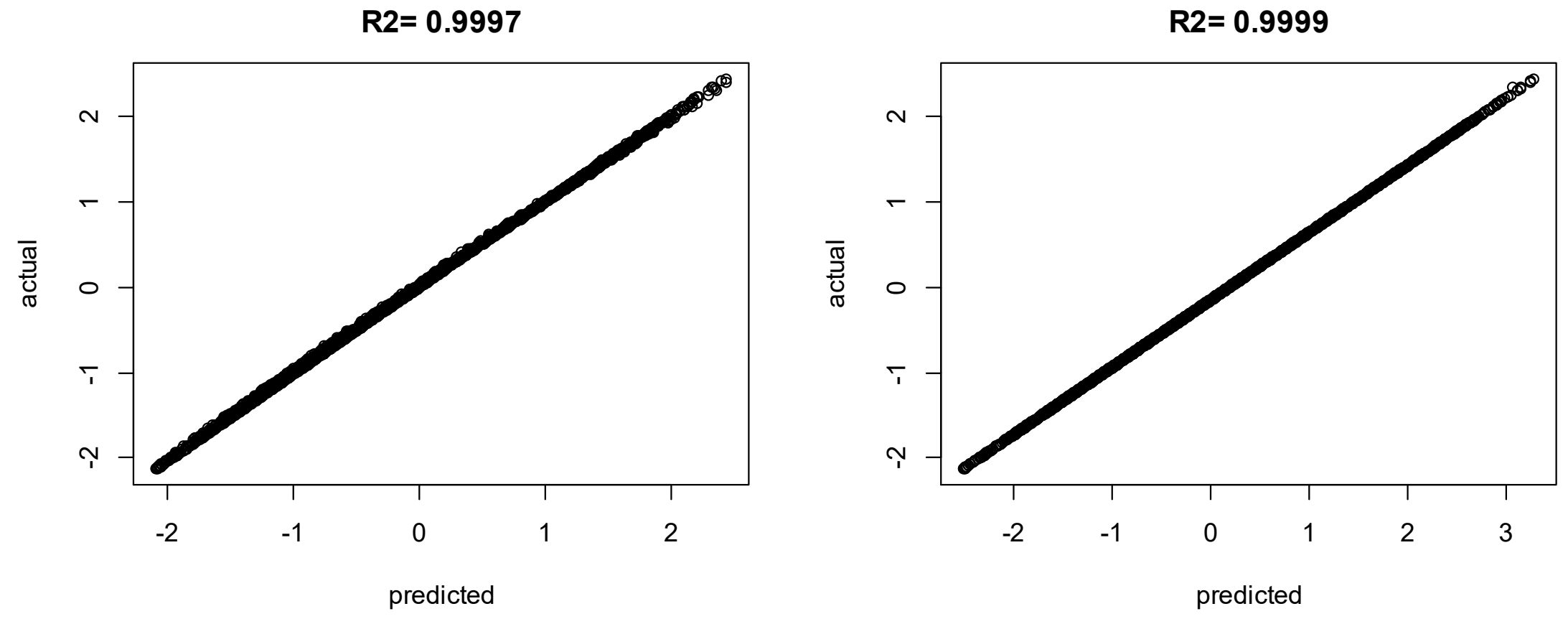

*Caption Figure 3: The plot on the left is the predicted vs observed value using the average linear fit to 26 dimensional delay maps with 25 independent dimensions predicting the third, extracting the 25 from a potential 30 dimensional delay map. The fit on the right is the average local linear fit using the same partition with a nearest neighbor size of 10*sqrt of n. The training is from the first 4000 steps in a 5000 step depiction of the Lorenz attractor.*

The results depicted above suggests **Conjecture 4:** for sufficiently high dimension, a linear model can perform similarly to the local linear model.

The total dimension being sampled from for the multiview prediction was 30.

**Application 2) Stock market**

In this case we attempted to predict three of the Dow Jones averages, the Dow Jones Industrial Average, the Dow Jones Transportation average and the Dow Jones Utility average. The data used for delay maps was the daily incremental change in the index value from market close to market close from 1948-2016. The delay maps were constructed reaching 80 time units back and the delay maps were of dimension 60. This was chosen through modeling the S&P 500 average over an 8 year time period. What was modeled was the 1$^{st}$ differences of each series, so the daily change in the Index

The actual correlation scatter plots are shown for each in figures 4, none of them seem to be interesting. But more interesting is whether the predictions could be used to instruct a sequence of decisions to trade in and out of the market, and beat the strategy of Buy and hold. Figure 11 shows the result of applying this to each index assuming a 3 point cost for each trade, (3/10ths of a percent of the value) and assuming no capital gains tax (red) vs a capital gains tax of 36% every two years

In figure 4, because there are approximately 17000 data points per cloud, the correlations may be taken to be representative for the ~70 year period, they don't appear to offer much hope for prediction. However below in figure 5 we can see that both the DJIA and the DJTA correlations allow us to choose whether to hold our position in the stock or either buy or sell all of it to change our position, in a fashion that in the long run over this ~70 year period does better than the buy and hold strategy, if we are not assessed capital gains. The method applied to the DJIA actually allows us to beat the buy and hold strategy even under a capital gains tax.

More importantly we can see that at least with respect to the Dow Jones industrial average, the collective action of humans on the stock market is predictable enough to make use of, so there is some hope of developing predictive models of social dynamics using nonlinear dynamic systems.

**Figure 4.**

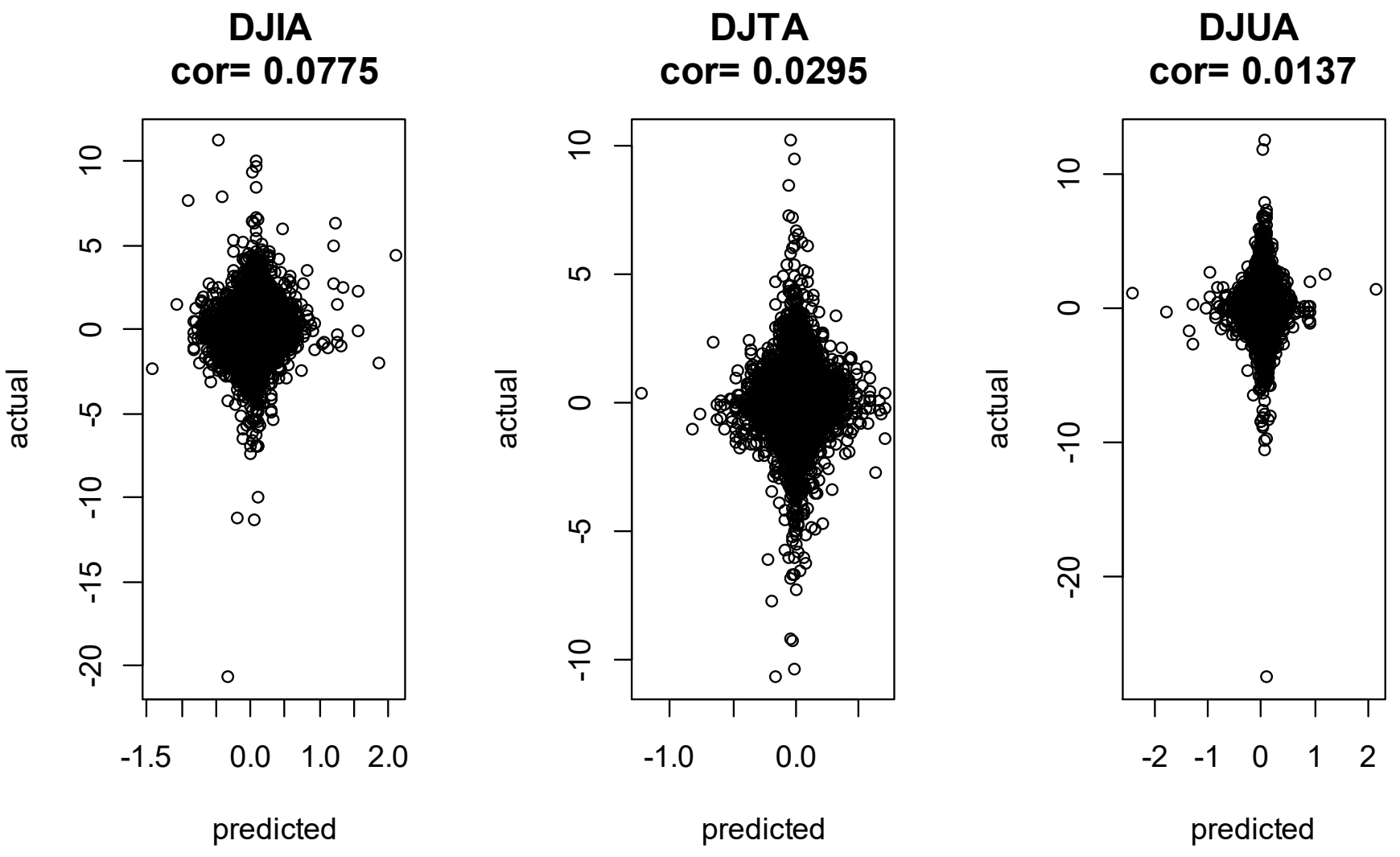

*Caption figure 4: Each plot shows predicted vs observed change for each day for the named Dow Jones average. The Correlations while small provide enough consistency for the Dow Jones Industrial Average and the Dow Jones Transportation average to "beat the market" by taking money out of the market when the predicted close tomorrow is below todays close and putting it in when the predicted close is above it.*

**Figure 5.**

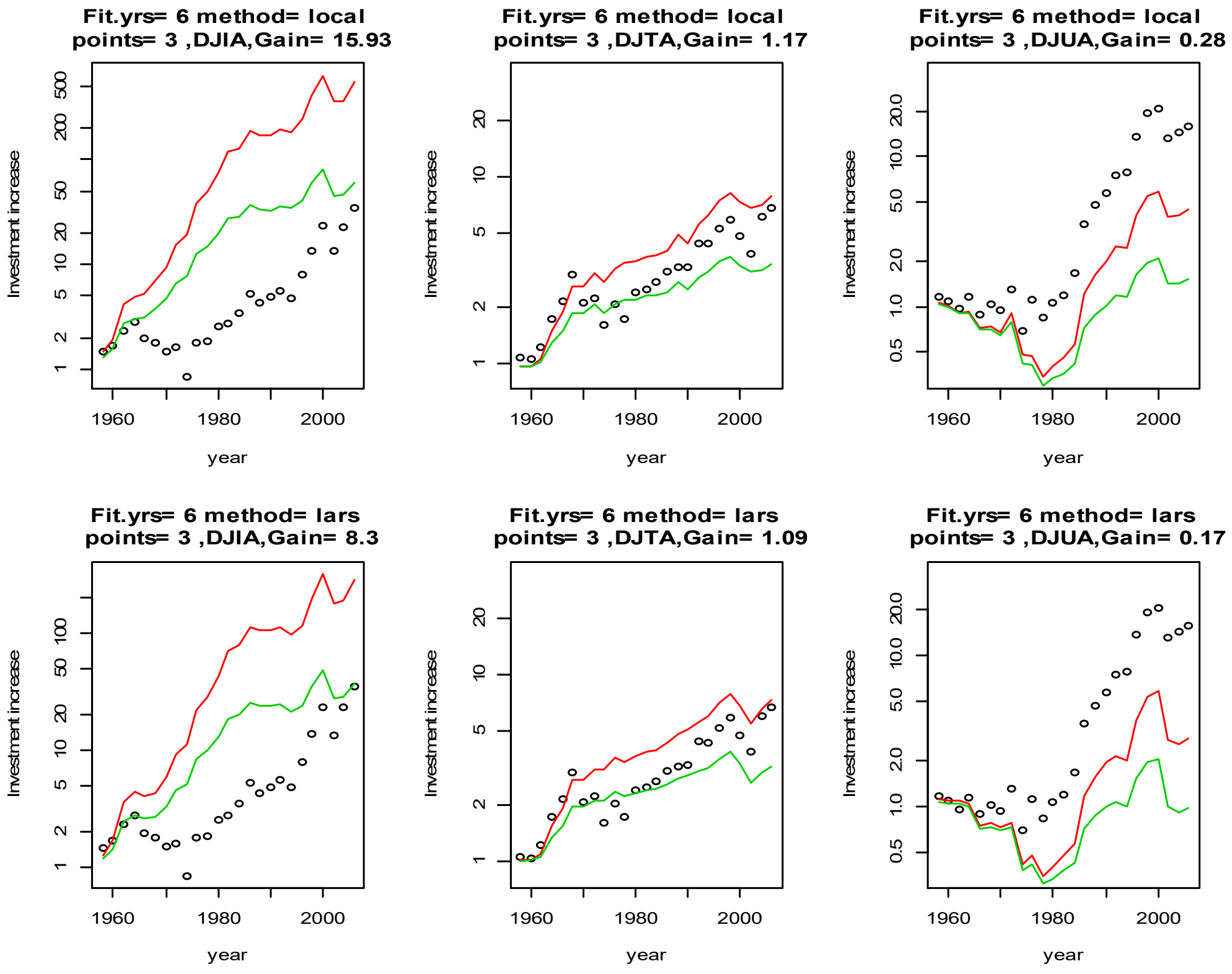

*Caption figure 5: The horizontal axis is year, the vertical axis is value of investment assuming $1 was invested in 1956. The top row uses the local regression method, the bottom a linear regression based on variable selection from the same partitions of delay maps using least angle regression. The black dots trace the buy and hold strategy every 2 years, the red lines are the stock market timing strategy with a 3 point cost per trade, and the green lines are the same strategy and cost except capital gains tax is extracted at 36% every two years.*

**Application 3) Precipitation in Volcano national park**.

In prior studies [7,8] it has been noted using mutiview embedding and Leith's approach to fluctuation dissipation that precipitation modeling based on using climate models to estimate the initial linear models and real data to tune the models allows fairly accurate predictions for a few seasons ahead (6 is the record). 10-15 year training periods of real data coupled with 100 years of computer model data were at a constant global average surface temperature were used. The skill of those predictions was in fact comparable to that achieved by the NOAA for the few regions over which this was done, only predicting further ahead.

The approach here, without using a computer based climate model was to take a 20 year time period, predict the last 2 years using the first 18, and find good parameter settings for p, window size, and percent of outlying predictions trimmed at each prediction point. Then the calculation was moved 2 years forward and repeated using the previously selected best prediction parameters. The result is shown for the later prediction. The prediction here is 1 year ahead, for monthly precipitation, with monthly mean precipitation subtracted (hence the negative numbers of the vertical and horizontal axis). Because the linear coefficients are shrunken in linear prediction for climate (probably because of the inherent measurement error in the X variable amplified by the exponential departure of nearby points over time) the predictions are rescaled to have the same range as the convolved induced SRB estimate (shortened to SRB estimate hereafter). Figure 12 shows the resulting predictions for both a linear response model and the local linear model. The SRB estimate from the training sample is overlaid in blue on top of the predictions.  The skill here is not overwhelming but still statistically significant.

**Figure 6.**

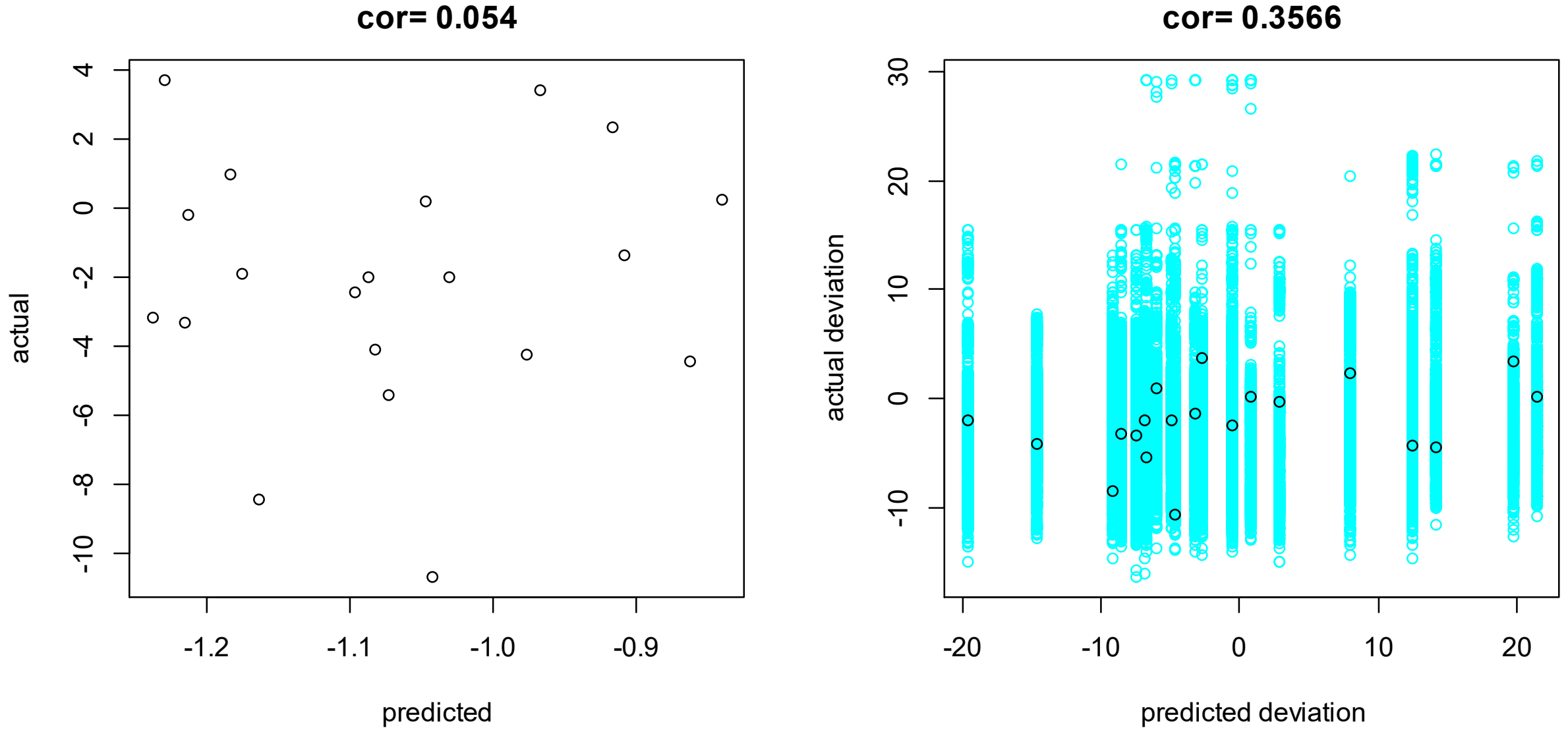

*Caption Figure 6: The plot on the left is the predicted vs observed value of monthly total precipitation using the average linear fit to the delay map for prior data from local precipitation, temperature, multivariate enso index , and pacific decadal oscillation index. The fit on the right represented by the black points is the average local linear fit using the same set of partitions. The blue are the residuals calculated as described from all the different fits representing the convolved SRB measure.*

Clearly empirical prediction intervals built on the SRB measures should contain the actual values. While it is not perfect prediction, the correlation (if we could think of the variables as normally distributed) would be better than chance. Applying several random partitions to create a distribution of these correlation estimates and then checking to see what the distribution looks like relative to a Gaussian distribution (with 125 recalculations using randomly selected partitions) we produce a quantile quantile plot shown in figure 7 which is consistent with a Gaussian distribution. Durbins version of the Kolmorgorov Smirnov  test[ref] for normality gives a p value of .71, assuming independence.

Increasing the number of points being predicted from 20 (as in figure 6) to 44 (using 40 years rather than 20 and predicting the last 10 %)  results in a change in the variance of the resampled  from .0249 to .0115 which gives a ration .0249/.0115 ~2.16, vs 44/20=2.2. So apparently each prediction point is acting as if it is uncorrelated. Making a leap of faith we could conjecture a rough 95% confidence bound for the correlation .355+/-.306 using the corresponding standard deviation of the “resampled” data.

**Figure 7.**

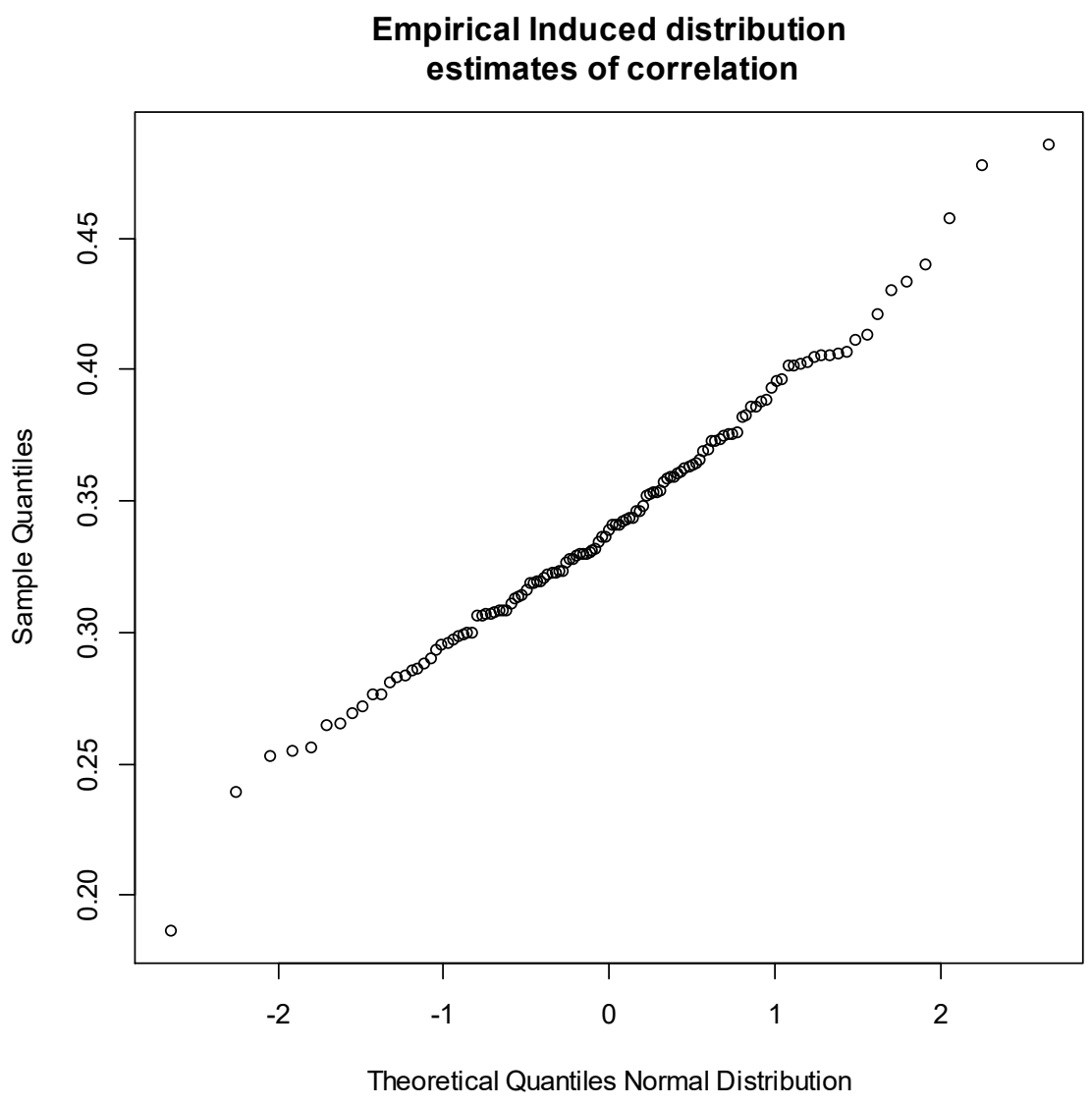

*Caption Figure 7: The points on the plot are correlations from 125 independent repetitions of the calculations on the right of figure 6, with independently defined collections of partitions.*

## Discussion and conclusion

In addition to the use of the SRB estimate as a way of providing prediction intervals, consider the following method of constructing conservative tests of goodness of fit of computer based models of dynamic systems.

Global climate models provide a method of estimating and predicting real climate events, however there is always a question as to how to determine, for example, if the parametrization of these models should be adjusted, and in what direction. Consider building a collection of predictions based on parameterizations in a neighborhood of a particular parametrization, and suppose that the parameterizations of those predictions can be considered to be in roughly a sphere surrounding a central point assumed to be the real parameters. Suppose we have R such parametrizations, uniformly distributed on the surface of the sphere. Suppose further that the attractor (and hence the SRB measure) is differentiable in all these parameters in that neighborhood. Then a test statistic, comparing a distance measure from a series of the actual residuals predicted in the real world from a combined distribution of predictions from the climate models. The differentiability means that the errors in the SRB measure should roughly cancel for the central point. Consider a simulation of this distance measure based on 1-fold cross validation of a set of similar residuals based on taking each of the R parameterizations in turn to produce pseudo observations vs the average of the rest of the models. The empirical distribution of the distance measure of the simulated residuals should be over dispersed with respect to an exact fit because of the variation added into the SRB measure by the parameter variation, so should allow conservative estimation of the probability of a distance measure as large or larger than that observed. Thus P values could be estimated for hypothesis tests. Further, the direction of each departure allows a clue as to what adjustment could be made to the parametrization.

The predictive distribution could also be used to modify the prediction. For example one could use modes from the distribution rather than directly using the regression determined estimate.

Using modes raises the question, could we for example guarantee the distribution would be unimodal, in which case the modes for each point would represent a prediction. An instructive example exists from an early attempt at prediction of regional precipitation in 2014. In this example, using a similar approach, a multimodal distribution appeared in a prediction for a number of stations in a region, the combined prediction is shown below. The Issue here is that there was a small very intense rainstorm (a “500 year” storm that passed through and hit one of the weather stations. The highest mode is over that storm, but it missed most of the weather stations. This did not include the residuals, but still illustrates a salient point, that there will still be some level of real uncertainty remaining in the system.

**Figure 8.**

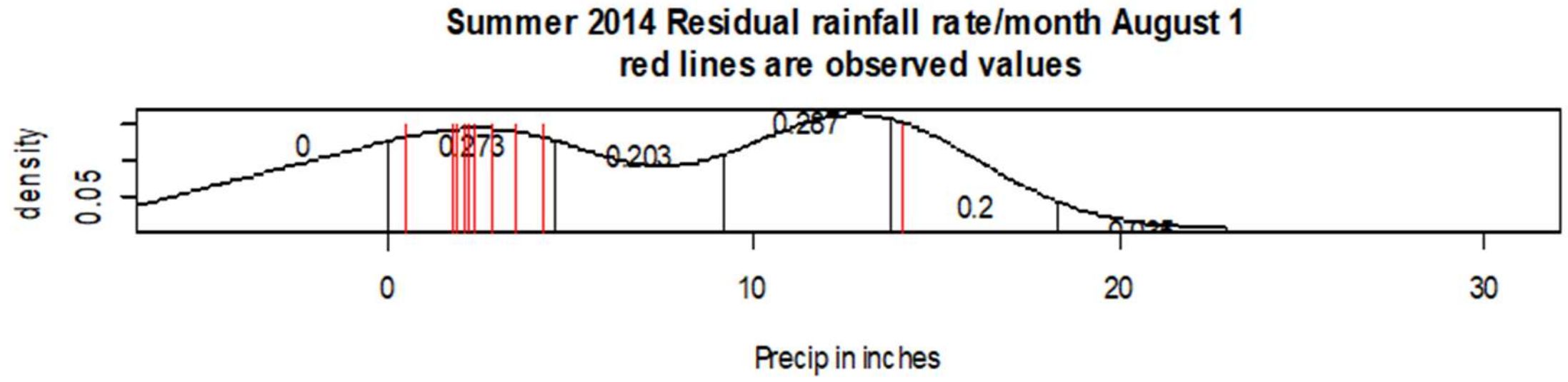

*Caption figure 8: In the summer of 2014 a distribution of precipitation predictions was constructed for a region containing a number of weather stations in the tristate area (New Jersey, New York, and Connecticut). Predictions were made for each weather station using each of a number of delay maps. The distribution was corrected at the beginning of August for the amount of rain that had already fallen in June and July, and the predictive distribution based on equally weighted least squares predictions for August is shown below along with the observed rainfall for august in red. The extreme red line is for Islip, New York where 14 inches of that are the result of one 24 hour period. The numbers in the smooth density sections are conditional probabilities assuming each set of p variables is equally likely to provide a useful prediction.*

*In this case the prediction distribution was not constructed from a history of the process, nor from nearest neighbor linear regression modeling. Instead it was constructed from regression models built around close geographical regions over 100s of years of climate models at constant greenhouse gas concentration, justified by fluctuation dissipation arguments (Leith[5]). Then subsets of the models were post selected for real data using an evolutionary algorithm (LuValle [7],[8]) to predict precipitation over earlier epochs of real time. So the density above is a density over data of the form* $\{w_{p_i}(x)H_{p_i}X_i\}_{i,p_i}$ *where* $p_i$ *is a selection of variables,* $w_{p_i}$ *is a weight determined by correlation in prior epochs,* $H_{p_i}$ *is the projection matrix for the regression and* $X_i$ *is the observed dependent delay map.*

Some theory has been proposed for building asymptotically consistent predictors for chaotic systems and a beginning has been made at looking at statistical inference based on the approach. No attempt has been made to suggest an optimal statistical approach, merely one that allows one to take the inherent uncertainty into account. However application of an appropriate modification of Bickel's approach may very well be optimal.

The software and data used in these analysis are available from the author at ml1305@stat.rutgers.edu. The stock market data [20] was obtained from Wharton Research Data Services. The data for volcano national park was obtained from the NOAA website [21], along with index data for the Multivariate Enso Index [22, 23], IOD[24], A0[25]. The PDO data was from the university of Washington web site [26].

With respect to the statistical climate modeling, the use of delay map embeddings as a basis for statistical analysis encourages the natural incorporation of teleconnections in ways that Markov random fields would not. I suspect it would also prove more useful in identifying subtle interconnections in other areas of application that are not immediately apparent.

A potentially interesting area for further mathematical study would be to follow up on the suggestion from the experiments with the Lorentz attractor that sufficiently high dimensional linear systems may be able to predict a chaotic system.

**Acknowledgments**

I would like to thank Michael Stein and William Strawderman for comments on an earlier draft.